# PAH chemistry at eV internal energies.

# 1. H-shifted isomers

Georges Trinquier, Aude Simon, Mathias Rapacioli, Florent Xavier Gadéa

Laboratoire de Chimie et Physique Quantiques (CNRS, UMR5626),
IRSAMC, Université Paul-Sabatier (Toulouse III), 31062 Toulouse Cedex, France

**Abstract**. The PAH family of organic compounds (polycyclic aromatic hydrocarbons), involved in several fields of chemistry, has received particular attention in astrochemistry, where their vibrational spectroscopy, thermodynamic, dynamic, and fragmentation properties are now abundantly documented. This survey aims at drawing trends for low spin-multiplicity surfaces of PAHs bearing internal energies in the range 1-10 eV. It addresses some typical alternatives to the ground-state regular structures of PAHs, making explicit possible intramolecular rearrangements leading to high-lying minima. These isomerisations should be taken into consideration when addressing PAH processing in astrophysical conditions. The first part of this double-entry study focuses on the hydrogen-shifted forms, which bear both a carbene center and a saturated carbon. It rests upon DFT calculations mainly performed on two emblematic PAH representatives, coronene and pyrene, in their neutral and mono- and multi-cationic states. Systematically searched for in neutral species, these H-shifted minima are lying 4-5 eV above the regular all-conjugated forms, and are separated by barriers of about one eV. General hydrogen-shifting is found to be easier for cationic species as the relative energies of their H-shifted minima are 1-1.5-eV lower than those for neutral species. As much as possible, classical knowledge and concepts of organic chemistry such as aromaticity and Clar's rules are invoked for result interpretation.

# 1. Introduction

Polycyclic aromatic hydrocarbons (PAHs) have received considerable interest since the initial proposal, in the mid-eighties, of their presence in the interstellar medium (Allamandola et al., 1985; Léger & Puget 1984). Astro-PAHs would be the carriers of the aromatic infrared bands (AIBs), a set of emission bands observed in the 3-14 μm range in regions where these molecules are heated by ultraviolet photons from stars. This proposal led to extensive experimental and theoretical studies to assign specific carriers, but so far no molecule has been unambiguously identified (Joblin & Tielens 2011). PAHs and their substituted derivatives also form a class of atmospherically relevant molecules because of their significant abundance, due to their efficient formation as byproducts of natural processes, such as biomass burning, or of human activities, such as combustion of fossil fuels (Finlayson-Pitts & Pitts 1987). On the other hand, PAHs constitute a well-documented class of organic molecules, the syntheses, structures and reactivity of which have been widely studied by organic chemists (Berionni et al., 2014; Bodwell 2014; Ciesielski et al., 2006; Hopf 2014; Müller et al. 1998; Popov & Boldyrev 2012; Scott 1982). If most of these studies limit to ground-state low-energy configurations and conformations - say within a few kcal/mol or 0.05 eV, there are situations where PAH molecules are rather hot, with internal energy much higher than this current organic-chemistry range as detailed hereafter.

In the interstellar medium, the PAH processing can be driven by interstellar radiation, stellar winds, shock waves or hot ionized gases (Micelotta et al., 2010a, 2010b). In photodissociation regions, at the border of molecular clouds where PAHs are exposed to star UV light, there is a competition between fragmentation and relaxation through radiative emission including IR emission (Allain et al., 1996a, 1996b; Bakes & Tielens 1994; Le Page et al., 2003; Montillaud et al., 2013). In the vicinity of stars, PAH may also collide with electrons (Siebenmorgen & Krugel 2010; Visser et al., 2007). The study of their energetic processing in earth laboratories has motivated experiments where they are excited by photons, often from synchrotrons since far-UV radiation is involved, (Boissel et al., 1997; Jochims et al., 1994; West et al., 2012, 2014; Zhen et al., 2015, 2016) or by collisions with high-energy ions (Champeaux et al., 2014; Lawicki et al., 2011; Martin et al., 2012;

Postma et al., 2010). These experiments have been complemented with theoretical studies (Chen et al. 2015; Holm et al. 2011; Jolibois et al., 2005; Paris et al., 2014; Postma et al., 2014). In these studies, PAH molecules are ionized and may remain rather hot after the excitation and ionization steps, possibly carrying internal energies of a few eV to tens of eV. Their relaxation dynamics is dominated by fragmentation with emission of H and $H_2$ as the lowest energy channels, although larger fragments such as $C_2H_2$ can be emitted as well.

To our knowledge, for commonly studied PAHs in astrophysical context, these lowest dissociation channels are lying, typically, around five eV above the ground-state in its equilibrium geometry. For these reasons, it appears desirable to explore the behavior of such molecules when they are subject to internal energies of a few eV, which clearly corresponds to energy conditions somewhat higher than they usually experience in common normal chemistry. The purpose of this work is to explore possible rearrangements along the wide landscapes of their potential-energy surfaces (PES) that are accessible within such energy ranges. The ensuing secondary minima should be considered in interstellar PAH processing studies, in particular regarding the search for new spectral features and PAH chemical evolution. To document these structural alternatives, we will try to keep advantage of the current chemical knowledge, as far as possible, and the set of simple concepts, laws and rules that contributes to "chemical intuition" will serve as a guidance all along our explorations.

Coronene and pyrene being emblematic PAHs and computationally tractable molecules, we will focus on these two systems, both in their neutral and cationic states. For these explorations, DFT (Density Functional Theory) is the method of choice, which will be used systematically all along this work. As PAHs are constituted of condensed benzenic units, an interesting question arising immediately concerns the π conjugation patterns. Here, simple approaches such as Clar's rules may shed light on structure and stability issues. Typically, they may account for how local alterations may have a global impact on π conjugation system and consequently on molecular geometry and stability. These arguments will thus be largely used all along the discussions.

The exploration is presented in two papers. In the present one, we examine how hydrogen atoms may undergo migrations to form H-shifted isomers. The H-shifted isomers happen to be precursors for various structural isomers such as vinylidene derivatives, as well as possible precursors for production of H or $H_2$ fragments. In the following companion paper, ring alteration and dissociation will be examined. The present paper is organized as follows. In Section 2, we supply computational details and briefly recall Clar's rules in PAH context. In Section 3, we use the results of quantum calculations to examine how hydrogen atoms may undergo migrations along the PAH carbon skeleton.

This work aims to bring out general trends for PAHs evolving on the surface associated to the ground state of their normal regular form. Therefore, this study will essentially limit to closed-shell *singlet* PES for neutral and dicationic species, and to *doublet* surfaces for monocationic and tricationic species, as these spin multiplicities correspond in general to the ground state of neutral and charged species in their *normal regular forms*. There is a notable exception to this inclination, namely the dicationic state of systems with degenerate highest-occupied orbitals, which may have a triplet ground state even in the regular form. This is the case for coronene, circum-coronene, *peri*-hexa-benzo-coronene, triphenylene, etc. (typically, the adiabatic energy difference $\Delta E_{T \to S}$ in dicationic coronene is calculated at 0.10 eV). Given that in H-shifted isomers, triplet states may also be favored for dications, this implies that for coronene dication, the triplet surface might lie below the singlet one *everywhere* in the portion of PES here explored. For neutral H-shifted isomers, the triplet state may also be lying below the singlet state, as it has been checked in several instances, but a systematic and complete study of both straddling states remains beyond the scope of the present work.

## 2. Methods and computational details.

Because they contain only carbon and hydrogen atoms, PAH molecules can be efficiently treated within DFT approaches. We used standard B3LYP hybrid functional with 6-311G** basis set. As mentioned above, some triplet states have been calculated for neutral and dicationic species at unrestricted UDFT level, as were treated mono-cationic and tri-cationic systems, expected to be monoradicals in doublet states as they carry odd numbers of electrons. In the latter cases, we have checked that $<S^2>$ (the expectation value of the $\hat{S}^2$ spin operator) is close to its anticipated value of 0.75. In rare specific cases, we got larger values ($<S^2> = 0.8 \sim 1.7$) pointing to spin contamination or contributions of doublet triradicals. In the same way, it might occur that an open-shell singlet diradical, with $<S^2>$ close to 1, be preferred to the expected closed-shell singlet configuration. These cases will be mentioned in due course. All calculations were performed using the *Gaussian09* quantum chemistry package (Frisch et al.). Full geometry optimizations were carried out up to energy gradients lower than $10^{-5}$ au. The natures of the so-obtained stationary points (minima or saddle-point) are determined through Hessian diagonalizations and vibrational analysis is systematically performed on these stationary points in order to characterize their nature. When given, frequencies correspond to uncorrected values. Likewise, no scaling factor is applied in the evaluation of zero-point vibrational energy (ZPE) corrections.

As PAHs are just assemblies of juxtaposed benzenic blocks, an important issue here is to gauge the extent of aromaticity carried by the whole system. In benzene, resonance between the two Kekulé forms leads to a symmetrical aromatic ring. In larger arrangements, a useful tool for global and synthetic analysis of π delocalization is given by the set of Clar's rules (Clar 1964, 1972; for recent developments on Clar's theory and applications, see for instance: Balaban & Klein 2009; Fujii & Enoki 2012; Maksic et al. 2006; Portella et al. 2005; Sola 2013; Wassmann et al., 2010). Briefly, they state that a maximum number of *disjoint* aromatic sextets (i.e. sextets separated by single bonds) will stabilize the molecule. The maximum-sextet picture is therefore supposed to best reflect both geometrical and electronic structures - and most often it actually does. As an illustration, in phenanthrene, pyrene, and triphenylene, there is a single Clar configuration obeying these rules (Scheme 1, top). In naphthalene and coronene (Scheme 1, bottom), two equivalent forms are possible and the resonance between them leads to higher symmetry of the whole structure, namely $D_{2h}$ for naphthalene and $D_{6h}$ for coronene. Molecular symmetry should actually reflect these effects, as aromatic rings are expected to exhibit identical bond lengths, hence higher symmetry, while purely single and double bonds are expected to exhibit larger and shorter bond lengths, respectively. One can legitimately wonder whether and to which extent these rules, which govern the lowest-energy structures of PAHs, will be helpful in the understanding of alternative local minima lying at several eV higher in energy, either for neutral or cationic states.

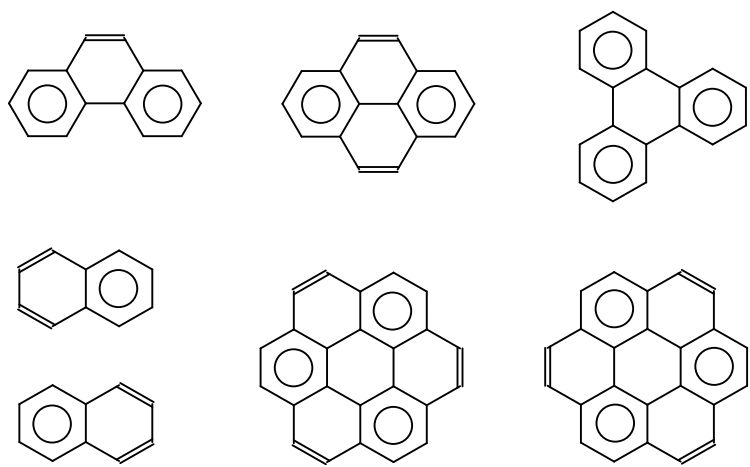

Scheme 1

**3. Hydrogen migrations: mechanisms and energetics.**

*A- Neutral species.* We first address hydrogen migration along the rim of a PAH. Such migrations leave one external carbon as a carbenic center, while creating, somewhere on outer rim or inner kernel positions an $sp^3$ saturated carbon center, either $CH_2$ or CH, depending if it affects an external 'convex' center, or an external 'concave' or internal carbon center respectively. 1,2-hydrogen shifts have been theoretically explored by Shaeffer *et al.* in organic chemistry (Nielsen et al., 1992; Schaeffer 1979; Yoshioka & Schaefer 1981), and were further readdressed by Jolibois *et al.* in the PAH context (Jolibois et al., 2005). Let us define, first, our conventions of notation. "1→2" will refer to the transfer of one hydrogen atom between two neighboring carbon atoms, "1→3" between two carbon atoms separated by a third one. On another hand, a "*m,n*" hydrogen-shifted isomer will define a structure in which carbon atoms *m* and *n* are under- and over-hydrogenated, respectively, with respect to their normal form.

For coronene and pyrene, we have explored all the pathway along the rim of the PAH. Due to the high symmetry of coronene in its equilibrium geometry, all external convex centers are equivalent, so that there is here only one kind of position for the carbenic center. Once formed this localized carbene or hydrogen 'hole', the symmetry is broken and the whole rim and inner positions need to be explored. For pyrene, three kinds of carbenic positions would need to be explored, two of them requiring the exploration along the entire outer rim and inner positions. For typical PAHs of decreasing symmetry, the total number of such isomers is detailed in Appendix A. Note that despite a lower number of atoms, there are more isomers for phenanthrene than for pyrene or coronene, owing to its lower symmetry. In pyrene and coronene, almost all of such H-shifted forms were found to be minima of rather low energy at physics eV scale, which is of course higher than the current chemistry kcal/mol scale. In some interstellar environments, hydrogen atoms of both pyrene and coronene have thus the possibility to migrate along the hydrocarbon rim, and one can conjecture this is a fundamental PAH property.

*Coronene*. For neutral coronene, there are seventeen possible local minima along the rim, as partly illustrated in Figure 1, top, with labeling convention from Scheme 2. Their relative energies with respect to the normal form are listed in Table 1, together with barriers linking consecutive minima. Not unexpectedly, due to steric hindrance, 'concave' and 'convex' sites behave rather differently, with a slight advantage to locate a $CH_2$ group at 'convex' sites rather than locating a CH group at 'concave' ones. Next, notice that the larger the distance between the hole and the saturated site, the higher the corresponding relative energy - a point somewhat supported by naïve intuition. Except for the 1-2 migration, all barriers are lying at about 1 eV above the related minima. Looking carefully at all these energies, one will notice a remaining pseudo symmetry along the circular pathway, reflecting the initial symmetry of the molecule.

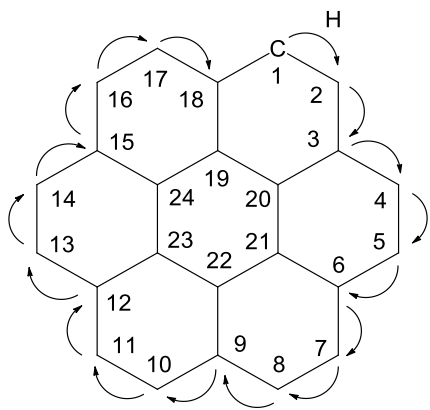

Scheme 2

Putting aside the first 1,2 isomer where saturation and hole are neighbors, all the 'convex' isomers are in the range 4.0 to 4.5 eV, while the 'concave' isomers are slightly higher, in the range 4.5 to 5.0 eV. The associated barriers range within 5.2 to 5.7 eV and 5.5 to 6.0 eV for the two kinds of corresponding barriers respectively. The energy landscape associated to the whole circular H-travelling along the molecular rim is plotted in Figure 2. The first 1,2-hydrogen shift, gives a -C-CH$_2$-pattern - a well-documented problem, for which barriers from singlet carbene forms range from 0.1 to 0.5 eV, depending on substitutents (Evansec & Houk 1990), and is associated with lower energies, either for the local minimum (3.51 eV), and for the corresponding barrier (3.70 eV). ZPE corrections lower both relative energies and barriers (see Table 1). The 1,18-hydrogen shift (still of 1→2 type), is lying at 4.43 eV, with a barrier to the most stable form at only 4.44 eV. After ZPE corrections, these two numbers decrease at 4.31 eV and 4.28 eV respectively, so that the barrier vanishes and the 1,18-hydrogen shifted isomer is no longer a real minimum.

All other cases correspond either to –H$_2$C–CH– → –HC–CH$_2$– (convex) or to –H$_2$C–C(sp$^2$)– → –HC–C(sp$^3$)H– (concave) migrations, outside local carbenic context. From the 1,2-hydrogen shifted minimum to the 1,3- one, the circular hydrogen travelling is broken as the system is inevitably caught into a broken-ring form bearing an extracyclic ethynyl group, and lying at only 2.71 eV above the normal form. The same transition state seems to link this low ethynyl minimum to both 1,2- and 1,3-hydrogen shifted isomers. These points should be shared by all PAHs, and, as ring alterations, they will be addressed subsequently in the second paper. Direct 1→3 transfers like –H$_2$C–C–CH– →

–HC–C–CH$_2$–, which avoids intermediate forms like –CH–CH–CH– are possible but are expected to be unfavored, except the saddle-point that relates the normal form to the 1,17-hydrogen shifted isomer, located at 5.06 eV (4.85 eV after ZPE corrections), which is slightly below the transition states linking 1,18- and 1,17-hydrogen-shifted isomers.

For the transition state relating 1,10 and 1,11 minima, beside the closed-shell solution at 5.74 eV, a slightly lower solution at 5.48 eV also exists, corresponding to a broken-symmetry open-shell singlet ($<S^2>$ =1.03), with a geometry quite close to the closed-shell one. This illustrates how high-spin solutions may become competitive when sites with hydrogen lack and excess are far apart. Concerning the saturations on the inner sites (labelled 19-24 in Scheme 2), they all appear to be real minima. Again, for sites near the hole (19 and 20), the energy is lower, comparable to the outer concave cases, while on the remaining sites (21-24) the energies are slightly above those of other positions. Reminding the above-discussed 1,2/1,3 situation, the saddle-point supposed to link 1,18 and 1,19 minima in fact relates both minima to an open form carrying a triple bond, located at only 4.07 eV (triply-bonded open forms are examined in the companion paper).

Besides these 1→2 hydrogen shifts where hydrogen connects sp$^2$ and sp$^3$ carbon centers, hole migrations should be considered as well, where hydrogen now connects two sp$^2$ carbon centers. We have explored some of these cases, in which the carbene hole is more or less far away from the CH$_2$ saturated site (See Table 2 and Figure 3, top). While intra-ring 1→2 hole migrations occur through out-of-plane hydrogen displacement, inter-ring 1→3 hole migrations occur through in-plane hydrogen displacement. In any case, these barriers are higher than those discussed above (6-7 eV *vs* 5-6 eV). Note that the two transition states determined for 1→2 hole migration correspond to open-shell singlet states ($<S^2>$=1.06 in both cases). With similar geometries, the closed-shell solutions lie about 1 eV higher in energy. Clearly, making the saturation move around is less energetically costly than making the hole move around - a result paralleled by imaginary frequencies, around 1250 cm$^{-1}$ for saturation migrations, while around 1750 cm$^{-1}$ for hole migrations, suggesting a more tightly bound transition state in the latter case.

To understand the labeling used in Figure 3, one must realize that 1→2 hole migrations in coronene convert a (1,*x*) isomer into its (1,21-*x*) counterpart, while 1→3 hole migrations convert a (1,*x*) isomer into its (1,20-*x*) counterpart. A global topological picture of our rim walk can be found as supplementary material (Figure S1), together with an example of the various ways of reaching a given hydrogen-shifted isomer (Figure S2).

*Pyrene*. In pyrene, there would be, *stricto sensu*, three kinds of circular walks for hydrogen migration, according to which one of the three possible starting points is chosen. Actually, we have explored the one in which the carbene (or hole) position preserves the Clar diphenyl-type frame (Scheme 3). Out of the 13 possible hydrogen-shifted minima partly illustrated in Figure 1, bottom, only 11 are found to be true minima, as 1,3 and 1,14 structures actually collapse into ethynyl and normal forms respectively, with a simple shouldering on the PES in the latter case. This leaves only two outer concave isomers, namely 1,7 and 1,10, and nine outer convex structures. The relative energies for the complete set of hydrogen-shifted isomers of neutral pyrene are listed in Table 3, and the energy profile along the outer rim walk is given in Figure 4. In this figure, pseudo symmetries remnant of the two mirror planes of $D_{2h}$ normal pyrene appear clearly at TS(8-9), and at minima 1,5 and 1,12. The 1,2-hydrogen-shifted isomer is largely stabilized as in coronene, but there is now a clear distinction among the other hydrogen-shifted isomers at convex positions. The most stable structures are 1,4, 1,6, 1,11 and 1,13 isomers, followed by 1,8 and 1,9 isomers, and nearly-degenerate 1,5 and 1,12 isomers. As in coronene, the concave minima (1,7 and 1,10) are the highest minima, lying here plainly above the inner minima 1,15 and 1,16. Interestingly, these trends are partly echoing simple topological Hückel arguments, as documented in Appendix B. When comparing Figures 2 and 4, pyrene exhibits lower energy minima than coronene, but most connecting relative barriers are comparable in both compounds.

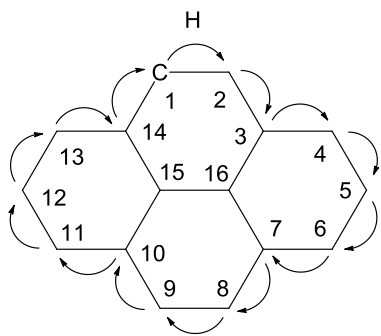

Scheme 3

Some examples of hole migrations have also been explored in pyrene (see Figure 3 and Table 2, bottom). We have not found here open-shell singlet states for 1→2 hole migration barriers. As in coronene, the hole barrier energies range within 6-7 eV, still with some advantage to in-plane 1→3 type over the out-of-plane 1→2 type. Note the lower value for the transition between 1,4 and 13,4 (5.8 eV). For both compounds, the barriers for *saturation migration* are around 1 eV, while those for *hole migration* are around 2 eV. Clearly, the saturation travels more easily along the rim than does the hole - a result that should be generalized to all PAHs.

A topological scheme linking the minima of Figure 4 along the pyrene rim walk is given as supplementary material (Figure S3). As mentioned, there would exist three kinds of such rim walks, depending on the choice for the position of the fixed outer carbene. Our choice implies that no pair of our hydrogen-shifted isomers can be connected through 1-3 hole migrations.

Relative energy profiles for all saturation migrations between neighbor carbons in the outer convex region are plotted in Figure 5. In coronene, where all such environments are virtually identical, the rearrangement is about isoenergetic and the barriers are fairly constant around one eV. In pyrene, the profiles depend on the concerned region. Not unexpectedly, the 8-9 path is comparable to the preceding cases since the environment is virtually similar to outer convex parts of coronene. In the 'sides' of pyrene, the profiles are asymmetrical. From 4 to 5 (or 11 to 12), the transfer is endothermic by about one eV, therefore concomitant with a higher barrier at 1.6-1.8 eV. From 5 to 6 (or 12 to 13), the transfer is exothermic by about one eV, therefore concomitant with a lower barrier at 0.7-0.9 eV.

While triplet states are higher than closed-shell singlet states in normal regular forms (at adiabatic ZPE level, $\Delta E_{S \to T}$ = 2.4 eV in coronene, and 2.0 eV in pyrene), in H-shifted isomers, the triplet state tends to be stabilized relatively to the singlet state. In 1,2-H-shifted isomers, both states are found to be exactly degenerate, as expected from an alkyl carbene. In 1,4 isomers and beyond, triplet states are lying below singlet states, with an associated gap $\Delta E_{T \to S}$ typically around 0.5 eV. Actually, the lower singlet state is here expected to be the diradical open-shell singlet counterpart lying just above and very close to the triplet state. A complete and systematic joint study of both surfaces would deserve separate publication.

*B- Charged species.* For mono-, di-, and tri-cationic species, only a selection of hydrogen-shifted isomers have been explored, including, for pyrene, some new types not included in the above discussion and drawn in Scheme 4. The corresponding relative energies and associated barriers are listed in Table 4. Upon single and triple ionization, there is a global trend to lowering both heights and barriers, but this occurs at different extent. The minima are lowered by one to two eV and the barriers by about one eV, the effect being more pronounced for coronene. Some of these charged minima are now lying as low as 2 eV above the normal form, and even lower than 2 eV for some

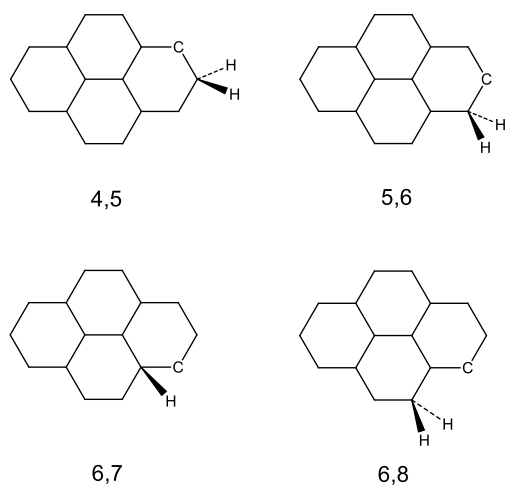

4,5     5,6

6,7     6,8

Scheme 4

pyrene trications. The 1→2 type hydrogen shifts corresponding to hole/saturation creation are further analyzed through averaged values (Appendix C) and illustrated by selected energy profiles (Appendix D). Di-ionization into singlet states perturbs less the profiles than mono- and tri-ionization into doublet states, but the effect is more conspicuous in pyrene. Table 4 indicates some discrimination between the three possible hole/saturation creations in pyrene. In neutral state, 1,2 is clearly preferred over 5,6, and 4,5, either from minima and barriers (see Table 4, bottom, left). In the monocation, the barriers still favor 1,2, although the minima would favor 5,6. All calculated barriers for monocations are larger than 3 eV, which is higher than that recently calculated for 1,2 hydrogen shift in naphthalene cation radical (2.77 eV) using a slightly different procedure (Solano & Mayer 2015). Actually, this apparent discrepancy does stem from the PAH nature and not from the basis set since from our procedure we get a similar value for this 1,2 barrier (2.79 eV, ZPE level), while the same 1→2 type hydrogen-shift barrier at 2,3 position is calculated at 3.07 eV - clearly, position matters.

Dicationic species can be preferentially triplet even in their normal forms, as soon as they have degenerate highest-occupied orbitals, as in coronene, where $\Delta E_{S \to T}$ = -0.10 eV, while in pyrene this gap is calculated at 0.67 eV. On the other hand, all hydrogen-shifted isomers at 1,2 position are expected to be triplet from the following reason. Upon creating *adjacent* carbene and $CH_2$ saturated centers, two (roughly) non-bonding orbitals are generated: the classical carbene $n_\sigma$ orbital, and a non-bonding π orbital, also localized, from topological arguments, on the *same* carbenic atom. Upon di-ionization, emptying these two non-bonding levels could seem natural, but this would locate two positive charges (Coulomb holes) on the same atom, a situation of course quite unfavorable. The system therefore prefers to leave an electron in the $n_\sigma$ orbital and to empty a lower π level in a triplet diradical configuration, with an open-shell singlet counterpart expected to lie close above in energy. This argument is strictly valid for 1,2-type H-shifted isomers. In this case, with respect to closed-shell singlets, the gap $\Delta E_{T \to S}$ are calculated at 0.55 eV in coronene, and 0.58 eV in pyrene (ZPE level). When carbene and saturated carbon centers are more distant, the triplet state seems to prevail for coronene, while it is not so for pyrene, where both states touch and cross. This can be seen in Table 4, where relative energies of triplet states are listed besides those of closed-shell singlets. From our sampling of di-cationic H-shifted minima (some of them are not given in Table 4), the triplet state is always preferred in coronene dication, which again arises the problem of the existence of a diradical singlet surface below the presently-discussed closed-shell one. By contrast, the triplet state is preferred in only seven cases over eleven in pyrene dication. Anyway, because this sampling is limited, a complete and systematic study of both surfaces would deserve to be carried out in a separate work. In the dication and in the trication, the 5,6 shift is clearly favored, while the 1,2 shift is the less favored one in the trication. As in neutral species, inclusion of ZPE lower both minima and barriers by 0.1 to 0.2 eV.

**4. Conclusion.**

By means of limited internal energies, PAHs can be converted rather easily from their regular all-conjugated forms into H-shifted isomers. As recapitulated in Figure 6, there are the three kinds of vicinal hydrogen shifts.

(i) The first H-shifted isomer, where carbene center and saturated carbon are neighbors, is lying at about 3.5 eV above the normal. The creation of this local minimum from normal forms requires rather large barriers, around 3-4 eV. Because of the endothermicity of the process, the reverse reaction to recover the *neutral* normal form requires a slight barrier of 0.2-0.3 eV. In cationic states, where the endothermicity is smaller, the barrier necessary to return to the normal form is higher, around one eV, thus giving some kinetic stability to the first hydrogen-shifted isomer.

(ii) Once formed, such isomer can propagate its saturation in other parts of the molecular frame through 1→2 hydrogen shifts of the *saturation migration* type. These require about one eV in coronene, and 0.5-2 eV in less symmetrical pyrene (see Figures 2 and 4). A hydrogen atom can therefore walk along the PAH rim and across the molecular plane, visiting H-shifted minima lying 4-5 eV above the normal form, and separated by barriers of about one eV. This process is made easier for cationic species as the relative energies of their H-shifted minima are 1-1.5-eV lower than those for neutral species.

(iii) The third kind of hydrogen migration relates a normal vertex –CH– to a carbenic one –C–, resulting in a 'hole' transfer. In neutral systems, such 1→2 or 1→3 hydrogen shifts were found to require barriers of 6-7 eV, which makes this process more energy demanding, and consequently less easy than the saturation migration.

In this paper, only the lowest spin-mutliplicity PES have been thoroughly investigated, while it was seen that in some cases higher spin-state minima become lower in energy. For neutral and dicationic PAHs, an exhaustive study of triplets and their associated open-shell diradical singlets (in such carbenic arrangements, they are expected to lie nearby in energy, and to mix in with closed-shell configurations) is welcome and should be the object of a dedicated work.

The present findings suggest that H-shifted isomers are rather universal as structural alternatives in the PAH world. As soon as a PAH bears a few eV of internal energy, H-shifting process may engage. In astrophysical context, such conditions are satisfied under UV irradiation or in collisions with protons, neutral hydrogen atoms or larger species, as well as in any combustion conditions. Therefore such a process should be included in PAH evolution modeling (spectroscopy, reactivity, fragmentation).

As discussed, the 'regular' walk can be interrupted by being trapped into catchment regions associated to opened ethynyl isomers. Such evolution calls for ring breaking, a more severe skeletal alteration, instances of which are examined in the forthcoming companion paper. Furthermore, H-shifting opens the way to scrambling, dissociation into H and $H_2$, and efficient conversion to vinylidene (or fulvene) structural isomers - all points carefully inspected in this second part. Finally, one could say that H-shifted isomers, ubiquitous in hot PAH chemistry, may be assigned to be possible links or bridges in a more general context of polymorphism.

**Acknowledgments** This work is part of the SWEET project (Stellar Wind and Electron interactions on astrophysical molecules. Experiment and Theory), integrated in the NEXT collaborative project, and involving teams of LCAR (CNRS-UMR5589), LPT (CNRS-UMR5152), and LCPQ (CNRS-UMR5626) laboratories.

Table 1. Relative energies of hydrogen-shifted local minima and barriers for neutral coronene.[a]

| saturated sites | | ΔE (eV) | |
| --- | --- | --- | --- |
| type | position | raw | ZPE |
| outer convex | 2 | 3.51 | 3.42 |
| | 4 | 4.28 | 4.18 |
| | 5 | 4.15 | 4.06 |
| | 7 | 4.45 | 4.35 |
| | 8 | 4.24 | 4.14 |
| | 10 | 4.45 | 4.35 |
| | 11 | 4.46 | 4.36 |
| | 13 | 4.14 | 4.05 |
| | 14 | 4.44 | 4.34 |
| | 16 | 4.13 | 4.04 |
| | 17 | 4.20 | 4.10 |
| outer concave | 3 | 4.58 | 4.48 |
| | 6 | 4.85 | 4.75 |
| | 9 | 5.02 | 4.92 |
| | 12 | 5.05 | 4.94 |
| | 15 | 4.63 | 4.54 |
| | 18 | 4.43 | 4.31 [e] |
| inner | 19 | 4.39 | 4.31 |
| | 20 | 4.68 | 4.60 |
| | 21 | 5.26 | 5.15 |
| | 22 | 5.31 | 5.20 |
| | 23 | 5.23 | 5.12 |
| | 24 | 5.16 | 5.05 |
| barriers | TS *n*-2 | 3.70 | 3.55 |
| | TS 2-3 [b] | | |
| | TS 3-4 | 5.52 | 5.30 |
| | TS 4-5 | 5.36 | 5.17 |
| | TS 5-6 | 5.69 | 5.48 |
| | TS 6-7 | 5.81 | 5.58 |
| | TS 7-8 | 5.62 | 5.41 |
| | TS 8-9 | 5.89 | 5.67 |
| | TS 9-10 | 5.99 | 5.76 |
| | TS 10-11 | 5.74 | 5.52 |
| | TS 10-11 [c] | 5.48 | 5.24 |
| | TS 11-12 | 5.98 | 5.75 |
| | TS 12-13 | 5.86 | 5.64 |
| | TS 13-14 | 5.59 | 5.39 |
| | TS 14-15 | 5.74 | 5.53 |
| | TS 15-16 | 5.63 | 5.43 |
| | TS 16-17 | 5.15 | 4.96 |

| | | |
|---|---|---|
| TS 17-18 | 5.12 | 4.93 |
| TS 18-*n* | 4.44 | 4.28 |
| TS 18-19 [d] | 4.79 | 4.66 |
| TS 17-*n* | 5.06 | 4.85 |
| TS 19-24 | 5.70 | 5.49 |
| TS 21-22 | 6.10 | 5.86 |

[a] Energy reference corresponds to the normal form; see Figure 1 and Scheme 2 for labeling; ZPE refers to relative energies taking into account zero-point vibrational energy corrections; TS stands for transition state; label *n* refers to the normal form.

[b] A structure of this type collapses into an ethynyl form, at 2.71 eV above the normal form.

[c] Broken-symmetry open-shell singlet state.

[d] This structure links the 19 minimum to the acetylenic form, at 4.07 eV above the normal form.

[e] At ZPE level, this minimum vanishes since it lies above the TS 18-*n* barrier.

Table 2. Calculated transition states for hydrogen shifts corresponding to hydrogen hole migrations (see Schemes 2 and 3 and Figure 3 for labeling).

|  | related minima | ΔE (eV) | |
| --- | --- | --- | --- |
|  |  | raw | ZPE |
| coronene | 1,5 - 1,16 | 6.94 | 6.67 |
|  | 1,10 - 1,11 | 7.22 | 6.94 |
|  | 1,5 - 1,16 (open-shell) | 6.27 | 6.00 |
|  | 1,10 - 1,11 (open-shell) | 6.36 | 6.08 |
|  | 1,4 - 1,14 | 6.44 | 6.20 |
|  | 1,8 - 1,10 | 6.43 | 6.19 |
| pyrene | 1,5 - 1,12 | 7.30 | 7.01 |
|  | 1,8 - 1,9 | 7.07 | 6.79 |
|  | 1,4 - 13,4 | 5.80 | 5.58 |
|  | 1,8 - 13,8 | 6.31 | 6.06 |

Table 3. Relative energies of hydrogen-shifted local minima for neutral pyrene, with corresponding barriers (TS).[a]

| saturated sites | | ΔE (eV) | |
| --- | --- | --- | --- |
| type | position | raw | ZPE |
| outer convex | 2 | 3.40 | 3.31 |
|  | 8 | 4.46 | 4.35 |
|  | 9 | 4.53 | 4.42 |
|  | 4 | 3.79 | 3.70 |
|  | 6 | 3.74 | 3.66 |
|  | 11 | 3.60 | 3.52 |
|  | 13 | 3.70 | 3.61 |
|  | 5 | 4.61 | 4.50 |
|  | 12 | 4.58 | 4.46 |
| outer concave | 3 [b] | | |
|  | 7 | 5.13 | 5.02 |
|  | 10 | 5.13 | 5.02 |
|  | 14 [c] | | |
| inner | 15 | 4.43 | 4.34 |
|  | 16 | 4.82 | 4.71 |
| barriers | TS *n*-2 | 3.60 | 3.45 |
|  | TS 3-4 [d] | 5.42 | 5.22 |
|  | TS 4-5 | 5.39 | 5.19 |
|  | TS 5-6 | 5.47 | 5.28 |
|  | TS 6-7 | 5.64 | 5.44 |
|  | TS 7-8 | 6.01 | 5.78 |
|  | TS 8-9 | 5.65 | 5.45 |
|  | TS 9-10 | 6.01 | 5.78 |
|  | TS 10-11 | 5.57 | 5.37 |
|  | TS 11-12 | 5.43 | 5.23 |
|  | TS 12-13 | 5.24 | 5.05 |
|  | TS 13-14 | 5.10 | 4.91 |
|  | TS 10-15 | 5.73 | 5.52 |

[a] Energy reference corresponds to the normal form (*n*); see Figure 1 and Scheme 3 for labeling.
[b] A structure of this type collapses into the ethynyl form, at 2.47 eV above the normal form.
[c] A structure of this type collapses into the normal form.
[d] Actually, this structure links the 1,4 minimum to the ethynyl form.

| | | neutral | | monocation | | dication [b] | | | | trication | |
|---|---|---|---|---|---|---|---|---|---|---|---|
| | | raw | ZPE | raw | ZPE | raw | | ZPE | | raw | ZPE |
| coronene | 1,11 | 4.46 | 4.36 | 2.41 | 2.42 | 2.43 | 2.22 | 2.35 | 2.18 | 2.35 | 2.34 |
| | 1,10 | 4.45 | 4.35 | 2.41 | 2.42 | 2.46 | 2.30 | 2.39 | 2.24 | 2.33 | 2.29 |
| | 1,18 | 4.43 | 4.31 | 3.03 | 3.05 | 3.32 | 2.90 | 3.26 | 2.87 | 3.18 | 3.17 |
| | 1,17 | 4.20 | 4.10 | 2.40 | 2.41 | 2.64 | 2.27 | 2.59 | 2.22 | 2.51 | 2.46 |
| | 1,2 | 3.51 | 3.42 | 2.35 | 2.35 | 2.75 | 2.17 | 2.68 | 2.13 | 2.29 | 2.29 |
| | *normal* | 0. | 0. | 0. | 0. | 0. | -0.11 | 0. | -0.10 | 0. | 0. |
| | TS (*n*-1,17) | 5.06 | 4.85 | | | | | | | | |
| | TS (1,17-1,18) | 5.12 | 4.93 | 3.62 | 3.54 | 4.16 | | 3.98 | | 3.58 | 3.50 |
| | TS (1,10-1,11) | 5.74 [c] | 5.52 | 3.26 | 3.18 | 3.14 | | 2.97 | | 2.94 | 2.84 |
| | TS (*n*-1,18) | 4.44 | 4.28 | 3.65 | 3.55 | 4.12 | | 3.96 | | 3.66 | 3.56 |
| | TS (*n*-1,2) | 3.70 | 3.55 | 3.26 | 3.17 | 3.70 | | 3.57 | | 3.06 | 2.96 |
| pyrene | 1,14 | | | 3.44 | 3.37 | 3.66 | 3.29 | 3.55 | 3.21 | 2.88 | 2.81 |
| | 6,7 | | | 3.44 | 3.36 | 3.98 | 3.31 | 3.86 | 3.22 | 2.75 | 2.71 |
| | 1,9 | 4.53 | 4.42 | 2.71 | 2.65 | 2.69 | 2.65 | 2.55 | 2.56 | 2.20 | 2.09 |
| | 1,8 | 4.46 | 4.35 | 2.73 | 2.66 | 2.83 | 2.64 | 2.70 | 2.55 | 2.22 | 2.15 |
| | 6,8 | 4.40 | 4.29 | 2.69 | 2.63 | 2.67 | 2.61 | 2.55 | 2.52 | 1.96 | 1.87 |
| | 4,5 | 3.90 | 3.78 | 2.87 | 2.80 | 3.34 | 2.59 | 3.22 | 2.50 | 1.87 | 1.84 |
| | 1,13 | 3.70 | 3.61 | 2.24 | 2.20 | 2.61 | 2.67 | 2.49 | 2.57 | 2.04 | 1.95 |
| | 5,6 | 3.69 | 3.59 | 2.20 | 2.15 | 2.24 | 2.60 | 2.16 | 2.52 | 2.10 | 2.00 |
| | 1,2 | 3.40 | 3.31 | 2.62 | 2.56 | 3.24 | 2.62 | 3.12 | 2.62 | 2.18 | 2. 13 |
| | *normal* | 0. | 0. | 0. | 0. | 0. | 0.71 | 0. | 0.67 | 0. | 0. |
| | TS (*n*-6,8) | 5.11 | 4.89 | | | | | | | | |
| | TS (6,7-6,8) | | | 3.93 | 3.77 | 4.50 | | 4.30 | | 3.04 | 2.92 |
| | TS (1,8-1,9) | 5.65 | 5.45 | 3.39 | 3.25 | 3.33 | | 3.13 | | 2.79 | 2.61 |
| | TS (*n*-4,5) | 3.99 | 3.83 | 3.51 | 3.36 | 4.09 | | 3.91 | | 2.75 | 2.61 |

| | | | | | | | | |
|---|---|---|---|---|---|---|---|---|
| TS (*n*-5,6) | 3.99 | 3.82 | 3.45 | 3.27 | 3.71 | 3.55 | 2.90 | 2.76 |
| TS (*n*-1,2) | 3.60 | 3.45 | 3.36 | 3.20 | 3.91 | 3.73 | 3.19 | 3.03 |

[a] In eV; same conventions as in Tables 1 and 3; see Figure 1 and Schemes 2 and 3 for labeling; empty box indicates no stationary point found.

[b] The second column refers to the triplet surface (no TS explored here).

[c] A singlet diradical form is found below this value, at 5.48 eV.

Table 4. Summary of relative energies for selected hydrogen-shifted cationic coronene and pyrene.[a]

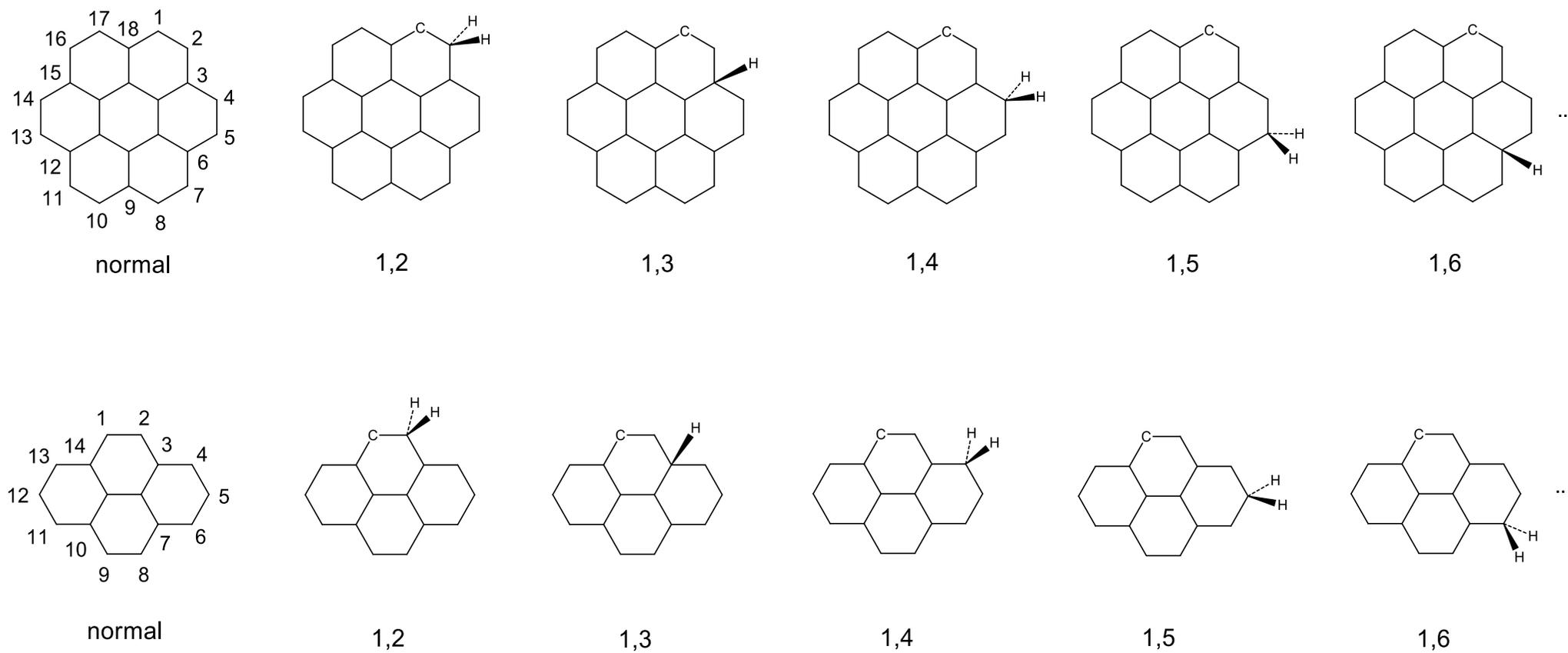

Figure 1. Principle of isomer labeling for hydrogen shifting along the outer rim of coronene (top) and pyrene (bottom).

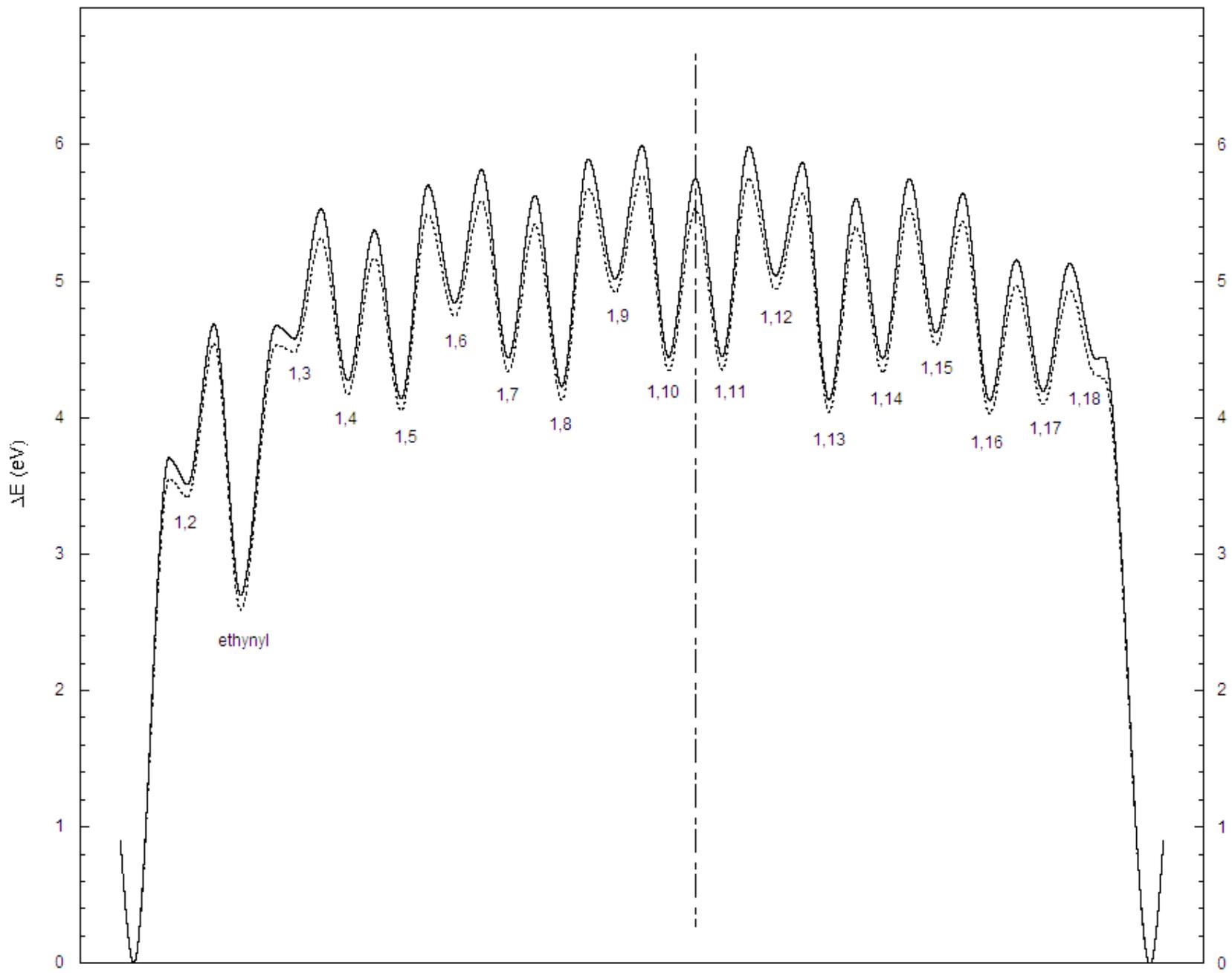

Figure 2. Energy profile for hydrogen migration along the coronene rim. See Figure 1 and Scheme 2 for the labeling of hydrogen-shifted minima. The dashed line takes into account ZPE corrections. The deep local minimum at left corresponds to an open ethynyl form - an unavoidable catchment region interrupting the circular travel.

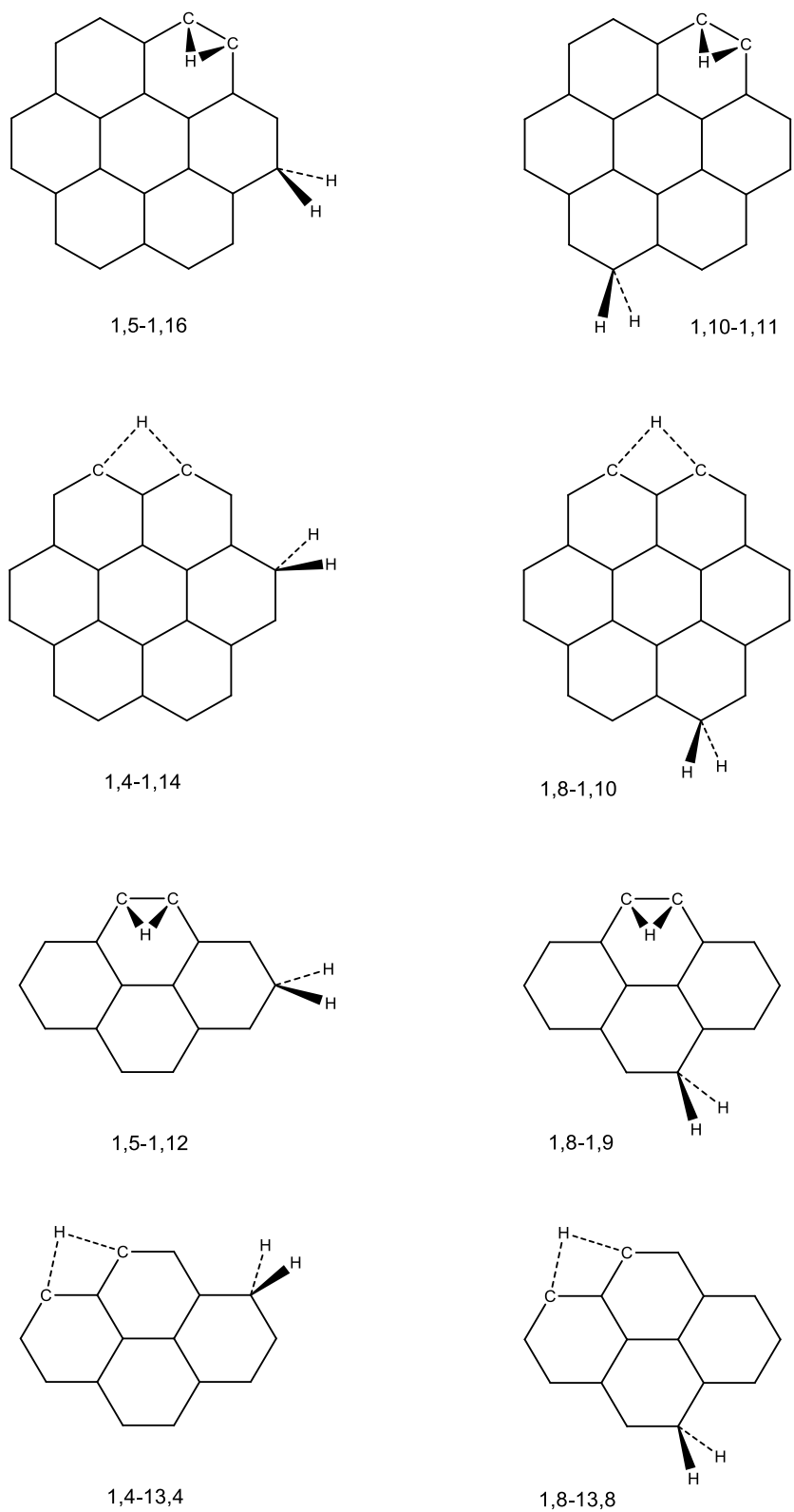

Figure 3. Examined cases of hole migrations of 1→2 type and 1→3 type in coronene (top) and pyrene (bottom).

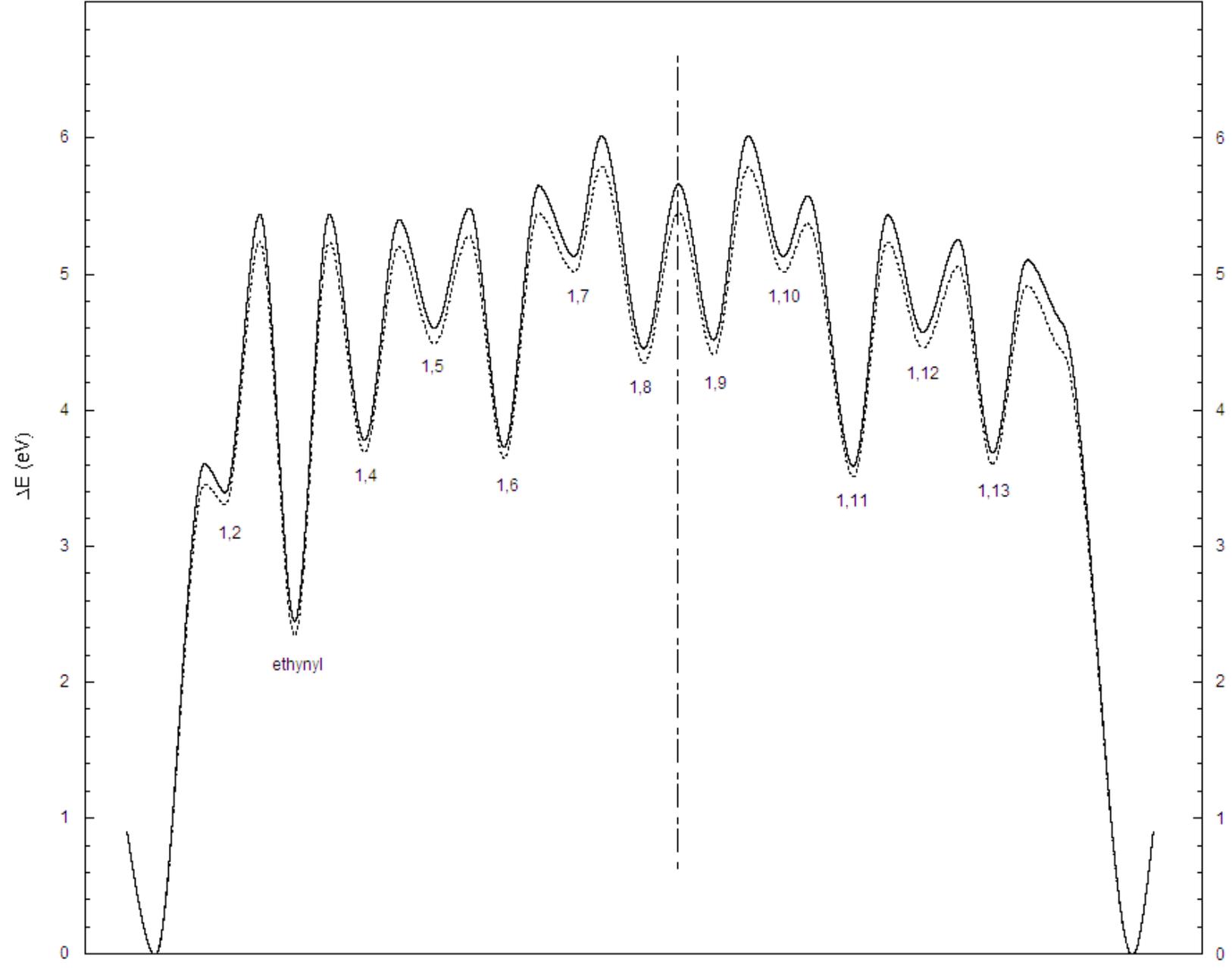

Figure 4. Energy profile for hydrogen migration along the pyrene rim. See Figure 1 and Scheme 3 for the labeling of hydrogen-shifted minima. The dashed line takes into account ZPE corrections. The deep local minimum at left corresponds to an open ethynyl form - an unavoidable catchment region interrupting the circular travel.

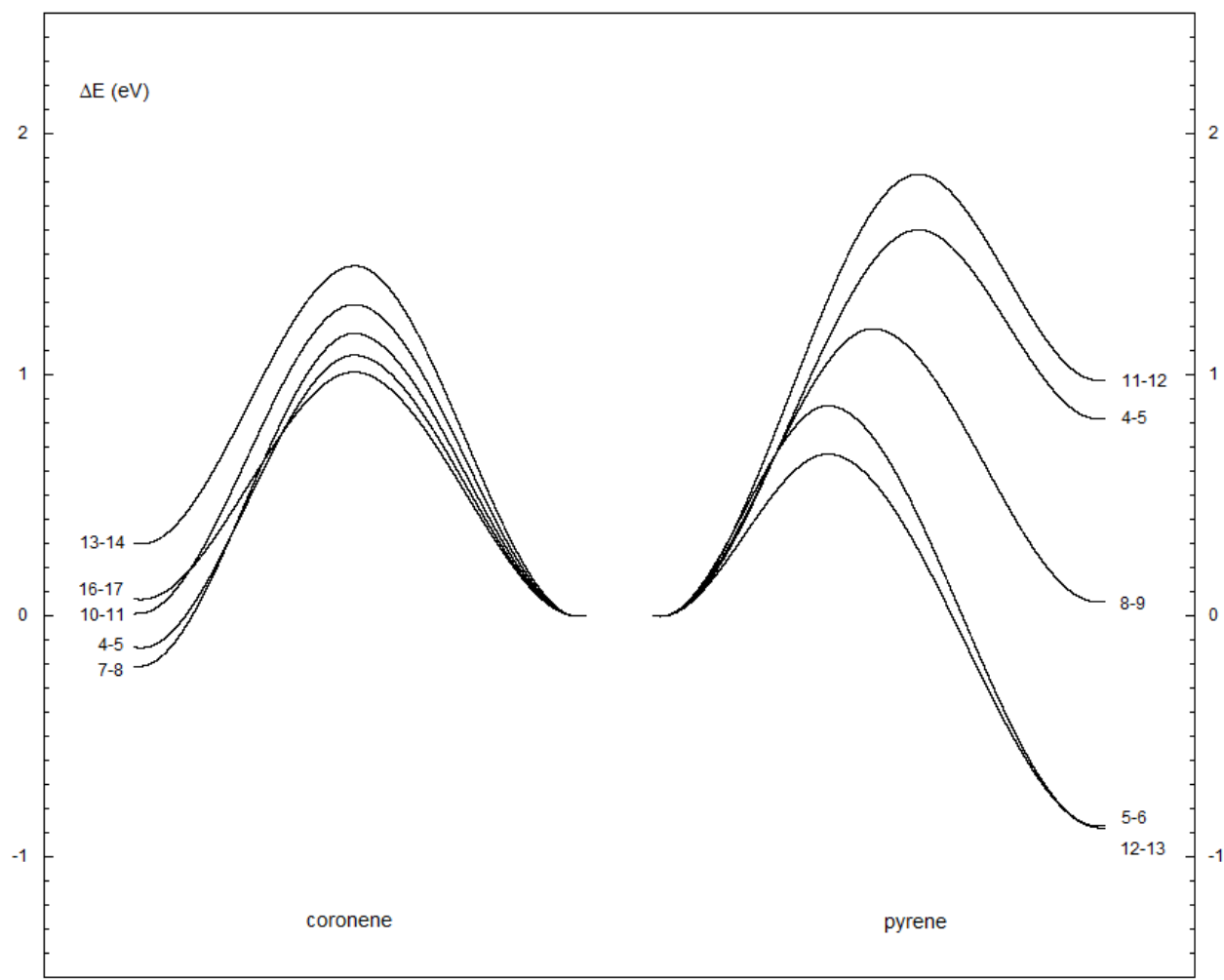

Figure 5. Schematic energy profiles for 1→2 hydrogen shifts corresponding to saturation migration H$_2$C–CH– → –HC–CH$_2$– along outer convex carbons of neutral coronene and pyrene; see Schemes 2 and 3 for labeling.

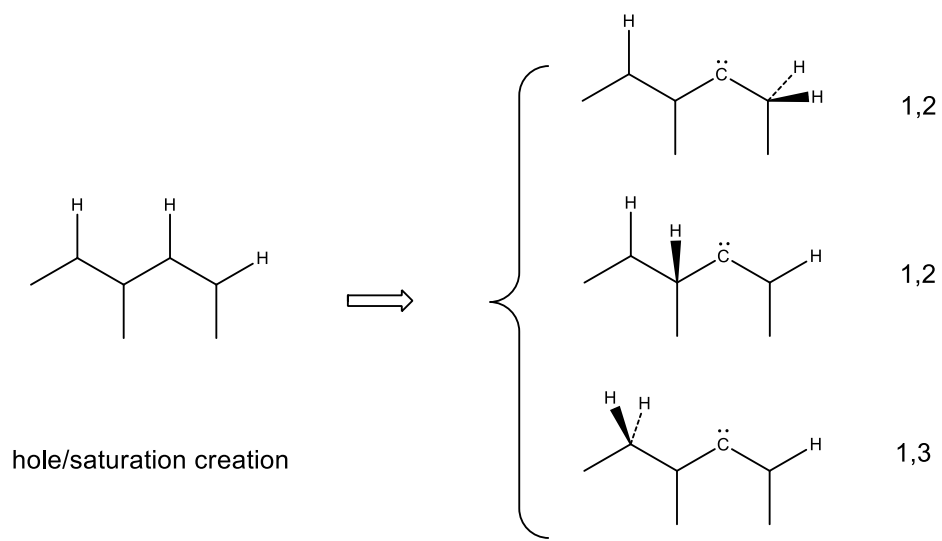

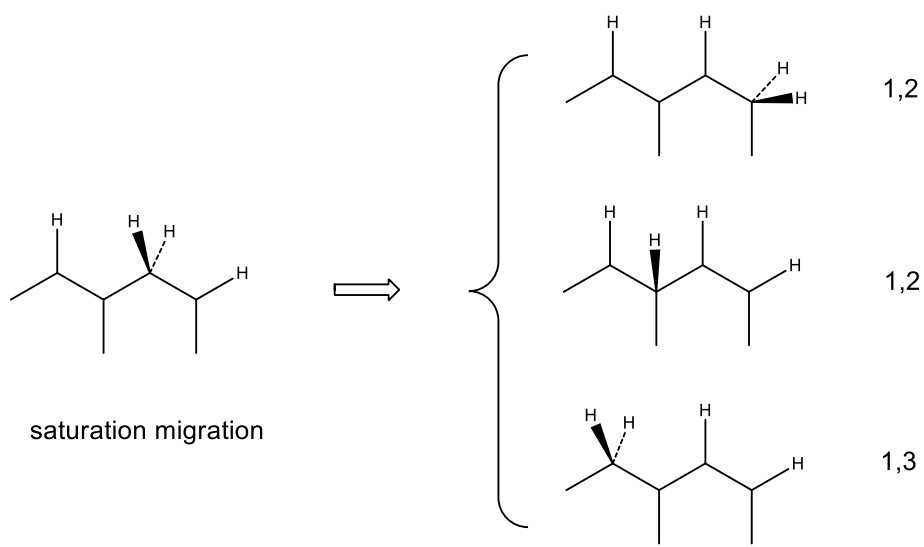

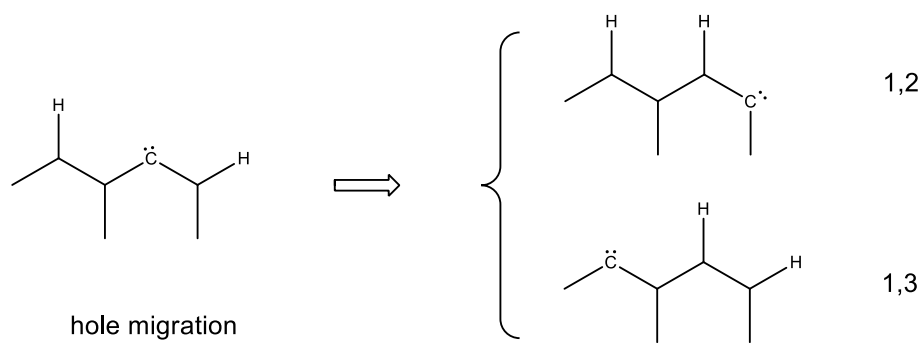

Figure 6. Various kinds of vicinal hydrogen shifts.

**Appendix A**. Hydrogen-shifted isomer counts.

|  | sym. | total | 1,2-convex | concave |
|---|---|---|---|---|
| coronene | $D_{6h}$ | 23 | 1 | 6 |
| pyrene | $D_{2h}$ | 39 | 3 | 10 |
| phenanthrene | $C_{2v}$ | 65 | 7 | 20 |

**Appendix B**. HMO relative loss of conjugation by selective saturation in coronene and pyrene ($|\beta|$ units).

|  | coronene | pyrene |
|---|---|---|
| **b** | 0.61 | 0.78 |
| **c** | 0.57 | 0.72 |
| $a_3$ | 0. | 0.36 |
| $a_1$ |  | 0.08 |
| $a_2$ |  | 0. |

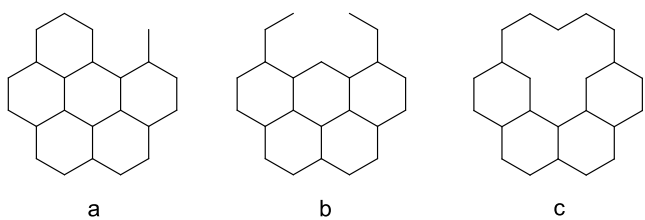

a            b            c

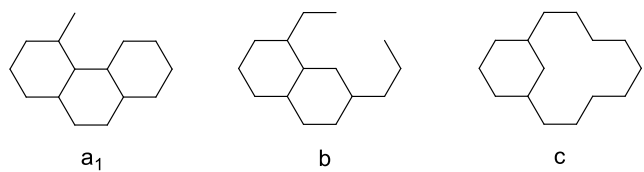

a₁           b            c

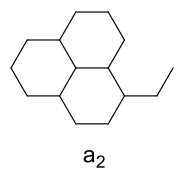

a₂

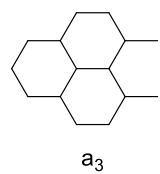

a₃

**Appendix C**. Mean values for 1→2 hydrogen shifts from normal forms. [a]

|  |  | raw | | ZPE | |
|---|---|---|---|---|---|
|  |  | mean | std. dev. | mean | std. dev. |
| neutral | $\Delta E^{\ddagger}$ | 3.82 | 0.20 | 3.66 | 0.19 |
|  | $\Delta E$ | 3.63 | 0.22 | 3.52 | 0.20 |
|  | $\Delta E^{\ddagger}_{\leftarrow}$ | 0.20 | 0.08 | 0.14 | 0.07 |
| monocation | $\Delta E^{\ddagger}$ | 3.40 | 0.11 | 3.25 | 0.08 |
|  | $\Delta E$ | 2.51 | 0.30 | 2.47 | 0.28 |
|  | $\Delta E^{\ddagger}_{\leftarrow}$ | 0.89 | 0.27 | 0.78 | 0.25 |
| dication | $\Delta E^{\ddagger}$ | 3.85 | 0.19 | 3.69 | 0.17 |
|  | $\Delta E$ | 2.89 | 0.50 | 2.80 | 0.48 |
|  | $\Delta E^{\ddagger}_{\leftarrow}$ | 0.96 | 0.36 | 0.90 | 0.35 |
| trication | $\Delta E^{\ddagger}$ | 2.97 | 0.19 | 2.84 | 0.19 |
|  | $\Delta E$ | 2.11 | 0.18 | 2.07 | 0.19 |
|  | $\Delta E^{\ddagger}_{\leftarrow}$ | 0.87 | 0.11 | 0.77 | 0.10 |

[a] In eV; averaged from migrations of 1,2-type for coronene and 1,2-, 4,5- and 5,6-type for pyrene. $\Delta E^{\ddagger}$ and $\Delta E^{\ddagger}_{\leftarrow}$ correspond to direct and reverse barriers, respectively.

**Appendix D**. Schematic energy profiles for 1→2 hydrogen shifts –HC–CH– → –C–CH$_2$– from normal coronene and pyrene in neutral (dashed curves) and ionized states.

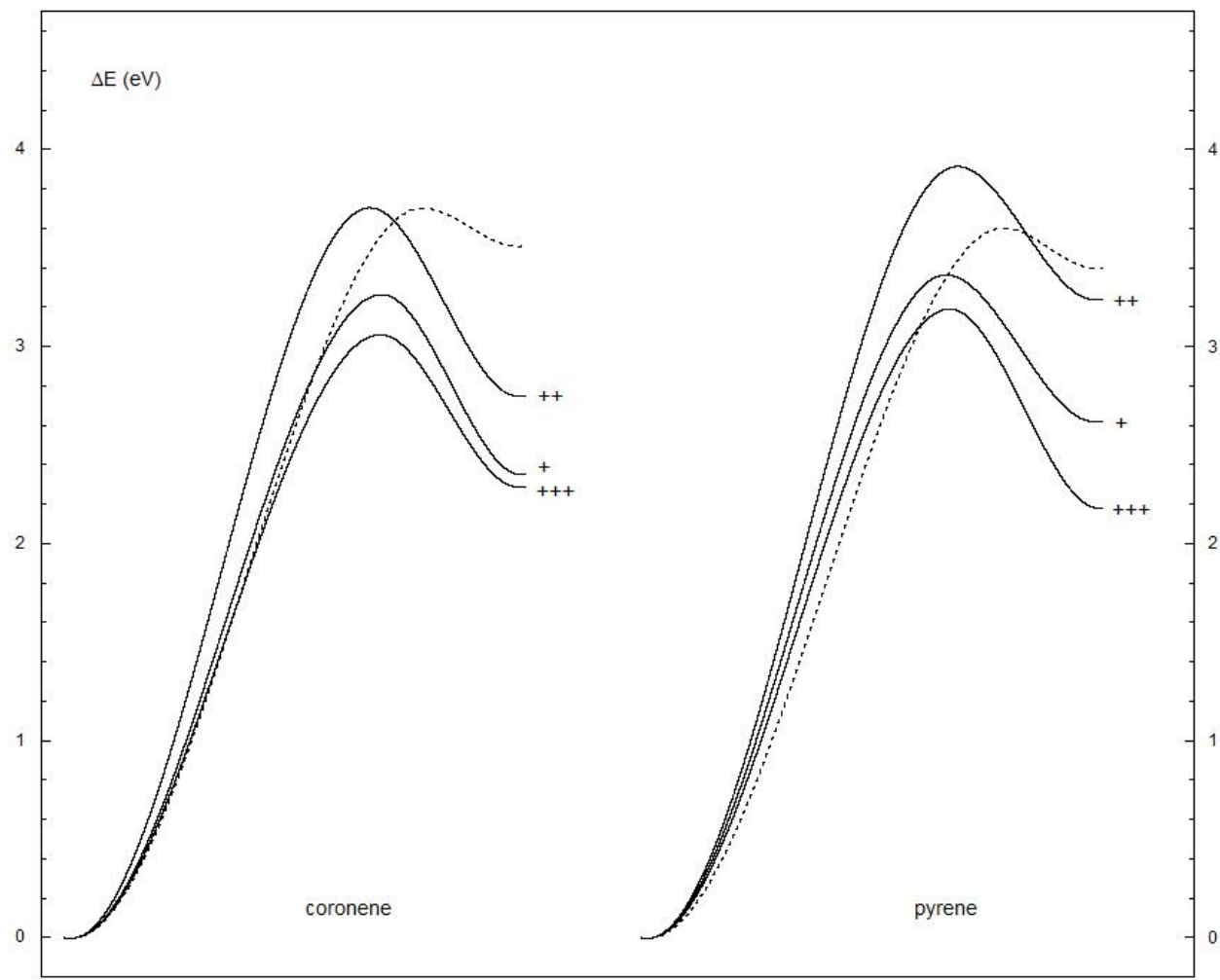